# Cross Site Request Forgery on Android WebView

Bhavani A B

Hyderabad, Andhra Pradesh 500050, India

**Abstract** - Android has always been about connectivity and providing great browsing experience. Web-based content can be embedded into the Android application using WebView. It is a User Interface component that displays webpages. It can either display a remote webpage or can also load static HTML data. This encompasses the functionality of a browser that can be integrated to application. WebView provides a number of APIs which enables the applications to interact with the web content inside WebView. In the current paper, Cross site request forgery or XSRF attack specific to android WebView is investigated. In XSRF attack, the trusts of a web application in its authenticated users is exploited by letting the attacker make arbitrary HTTP requests on behalf of a victim user. When the user is logged into the trusted site through the WebView, the site authenticates the WebView and not application. The application can launch attacks on the behalf of user with the APIs of Webview, exploiting user's credentials resulting in Cross site request forgery. Attacks can also be launched by setting cookies as HTTP headers and making malicious HTTP Request on behalf of victim.

**Keywords** - *Android WebView, Cross Site Request Forgery, Computer Security.*

## 1. Introduction

The Android platform's openness and permissive licensing allows the software to be freely modified and distributed by device manufacturers, wireless carriers and developers. As a result of this, Android has a large community of developers writing applications. The number of applications in the app store has been estimated to be 1 million with about 50 billion downloads till date [18]. A report in July 2013 stated that Android's share of the global smart phone market was 64% in March 2013 [19]. The popularity of Android platform has led to a large number of third party applications in the app store which can be a potential cause of security threats. Many of the Android applications display web content and also interact with it. This is possible by exposing a web browser as a standalone component and embedding it in the application. Such a component is called as WebView. Android comes with a build-in high-performance WebKit browser and applications can access views, that render using WebKit through WebView component.

### 1.1 Introduction to Web-based APIs of WebView

There are two types of APIs in WebView, the UI based APIs and Web-based APIs. WebView is subclass of a more generic View class. View occupies a rectangular area on the screen and is responsible for drawing and event handling. View is the base class for widgets, which are used to create interactive user interface components such as buttons, text fields, etc. The APIs which the WebView inherits from its superclass are UI based APIs. Attacks described in [1] target the UI based APIs. Web-based APIs are implemented by the classes associated with WebView. These APIs are designed for applications to interact with the web contents [1]. Examples of Web-based APIs include loadUrl, loadData etc. The attacks described in [2] target Web-based APIs. These include Event sniffing, hijacking, JavaScript injection and attacks through frame confusion.

Cross-site scripting attacks using Web-based APIs are described in [3] and [12]. Web-based APIs of WebView are discussed in detail in section 2.

### 1.2 Overview of Work and Contribution

The present paper describes Cross Site Request forgery on Android WebView. When the user is logged into the trusted site using his login information, the target site authenticates the WebView and not the application. The application can exploit user's authenticated session and launch attacks, resulting in Cross site request forgery. The attacks can be modeled using three of the Web-based APIs, WebView.loadUrl, WebView.loadData and WebView.postUrl. Such attacks result in automatic form submissions with user's credentials. Attacks can also be modeled by sending malicious HTTP requests to the server exploiting user's credentials. The cookies from WebView can be gathered from the CookieManager.getCookie() method, and can be set as HTTP headers. The HTTP request will now contain user's login information such as session ID that is gathered from cookies. Attacker can send malicious HTTP GET or POST to the trusted site, with user's login information. HTTPClient APIs are described in detail in Section 3.

## 2. WebView APIs

WebKit is a layout engine for a large number of browsers including Google's chrome web browser and default browser in Android. The android.webkit package provides tools for browsing the web. WebView is one of the most important classes of the package. It enables the developer to embed a built-in Web browser as widget, for displaying HTML content and browsing the web. In addition, WebKit is a layout engine designed to allow web browsers to render web pages. It serves as a backend rendering to WebView. The android.webkit provides several other classes such as CookieManager, CookieSyncManager, WebSettings, WebStorage etc. Jointly, these classes expose many APIs to Android applications. Based on their purposes, these APIs can be divided into two main categories, the Web-based APIs and UI based APIs [2]. The current paper describes attacks related to Web-based APIs of WebView and also related to the CookieManager class.

### 2.1 CookieManager

A cookie, also known as an HTTP cookie, is a small piece of data sent from a website and stored in a user's web browser while a user is browsing a website. The data stored in the cookie is then sent back to the server each time the browser requests a page from the server. The CookieManager class in Android manages cookies used by an application's WebView. This class provides a number of public methods such as getCookie, acceptCookie() etc.

## 3. HttpClient APIs

The Hyper-Text Transfer Protocol (HTTP) is perhaps the most significant protocol used on the Internet today. HttpClient provides an efficient, up-to-date, and feature-rich package implementing the client side of the most recent HTTP standards and recommendations. HttpClient Class is an interface for HTTP client. HTTP clients encapsulate objects required to execute HTTP requests while handling cookies, authentication, connection management, and other features. The most essential function of HttpClient is to execute HTTP methods [23]. Execution of an HTTP method involves one or several HTTP request or HTTP response exchanges, usually handled internally by HttpClient. The user is expected to provide a request object to execute and HttpClient is expected to transmit the request to the target server to return a corresponding response object, or throw an exception if execution was unsuccessful. Here is an example of request execution process:

HttpClient httpClient = new DefaultHttpClient();

As shown in the code above, DefaultHttpClient is the default implementation of the HttpClient interface[20].

### 3.1 HttpRequest

HttpClient supports HTTP methods defined in the HTTP/1.1 specification [10], GET, HEAD, POST, PUT, DELETE, TRACE and OPTIONS. There is a specific class for each method type: HttpGet, HttpHead, HttpPost, HttpPut, HttpDelete, HttpTrace, and HttpOptions. The Request-URI are a Uniform Resource Identifier that identifies the resource upon which to apply the request. HTTP request URIs consists of a protocol scheme, hostname, optional port, resource path, optional query, and optional fragment. The code fragment is given below [24]:

HttpPost httpPost = new HttpPost(
                "http://www.targetSite.com/cgi-bin
                     /Forum/new_pm.php");

### 3.2 HttpResponse

HTTP response represents the message returned to a client in response to an HTTP request. The first line of that message consists of the protocol version followed by a numeric status code and its associated textual phrase. The execute method executes the request using the default context and returns the response to the request [24].
HttpResponse response =
httpClient.execute(httpPost);

### 3.3 HTTP Message Headers

A HTTP message can contain a number of headers describing properties of the message such as the content length, content type and so on. The method setHeader overwrites the first header with the same name. The new header will be appended to the end of the list, if no header with the given name can be found. The method takes two arguments, first one being the name of the header and second one being the value of the header [25].

httpPost.setHeader("Cookie", cookie);

# 4. Attack Model

For all the attacks described in the paper, we have the following assumptions:

1) User must be authenticated to the trusted site: Cross site, Request forgery attacks exploit the user's credentials after the user is logged into the trusted site throughWebView. Therefore an assumption is being made that the user must be logged into trusted site, for the attacks to be successful.

2) Application must be granted permission to access the Internet: For the attacks described throughout the paper, the application needs to be granted with permission Android.permission.INTERNET. This permission is granted to 65 percent of the paid applications and 86.6% of free applications [4].The attacks are quite easy to launch as these permissions are granted to most of the applications.

3) Attacks are related to third party applications: The paper describes the vulnerabilities in the third party Android applications. The developer of the third party application and owner of web content inside WebView are not the same. Therefore, there is a potential threat from a malicious application installed on the android device.

# 5. Cross Site Request Forgery

The Cross site request forgery is an attack which forces end user to execute unwanted actions on a web application using his authority or credentials. It is one of the top 10 security vulnerabilities of 2013 [6]. In a CSRF attack, a malicious site instructs a victim's browser to send a request to honest site, as if the request were part of the victim's interaction with the honest site, leveraging the victim's network connectivity and the browsers state, such as cookies, to disrupt the integrity of the victim's session with the honest site [7]. Depending on the application, the attacker could, for instance, post messages in the name of victim, or even change victim's login name and password. XSRF attack is much different from the XSS or Cross site Scripting attacks described in [3]. In XSS attacks, the attacker executes a malicious script on the victim's browser and steals sensitive information such as cookies. In XSRF attacks, the attacker sends malicious requests on victim's behalf by exploiting the authenticated session of victim. Unlike cross-site scripting (XSS), which exploits the trust a user, has for a particular site, XSRF exploits the trust that a site has in a user's browser or WebView. XSS is about injecting malicious scripts in web pages. The malicious scripts in turn gains access to sensitive information and begin to misuse it. In XSRF, the attacker exploits the credentials of the user who is authenticated to the trusted site and sends HTTP requests to that site on behalf of the user.

In the current paper, Cross site request forgery specific to android WebView is investigated. When the user is logged into trusted site, the site authenticates WebView and not the application. The application can exploit the user's session and launch attacks on the behalf of user with the APIs of WebView. The attacks are modeled using three of the WebViews APIs, loadUrl, postUrl and loadData. The attacks can also be modeled by the HTTP GET or HTTP POST requests. When the user is authenticated with trusted site, cookies can be gathered through the CookieManager. The cookies which are gathered can be appended to the list of headers of the HTTP Request using the method setHeader. These attacks are described in detail in the following subsections.

## 5.1 Attacks using HttpRequests through WebView.loadUrl

One of the methods to perform POST/GET requests to another host is to use the HTML form tag with the action attribute set to the target URL. The malicious HTML file is provided as a part of the application in the assets folder in Android device. The HTML form is shown in Fig 1. The required form fields are set the desired values, in the HTML file. The attacker can now execute the HTML file, and submit the form programmatically by using webView.loadUrl.

webView.loadUrl("http://trustedSite.com/ cgi-bin/Forum/index.php");

```
<html>
    <head>
    </head>
    <body>
        <form id="post-form" action="http://www.targetSite.com/cgi-
bin/Forum/new_pm.php" method="post">
            <input type="hidden" value="WebView Attack from android"
id="title" name="title" /><br />
            <input type="hidden" value="sohini" id="recip" name="recip" />
            <input type="hidden" value="WebView attack message from
Android" id="message" name="message" />
            <input type="submit" value="Send" />
        </form>
        <script type="text/javascript">
        document.getElementById("post-form").submit();
        </script>
    </body>
</html>
```

Fig. 1 Malicious HTML form

When the user is logged into the target site using his credentials, the site authenticates WebView, and not the application. The malicious application can now exploit the user's credentials and submit POST requests on behalf of the user, resulting in the XSRF attacks, as shown in Fig. 2. The attacks can also be modeled using GET requests.

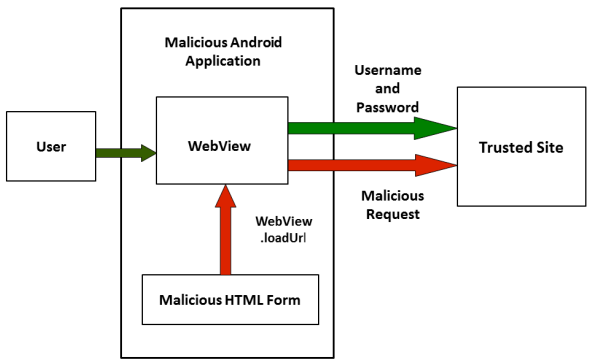

Fig. 2 Attacks using Malicious HTML form.

## 5.2 Attack using HTTP requests through WebView.loadData

WebView loads raw HTML data using loadData method. The method takes three input parameters, a string of data in the given encoding, MIME type of data and the encoding of data, which specifies whether it is base64 or URL encoded. The HTML data containing GET or POST request is assigned to a string, and passed as a parameter to the WebView.loadData. The malicious requests to the server through the API, can leverage the user's login session, and submit HTML form automatically on behalf of user.

```
webView.loadData(attackData, "text/html;
charset=utf-8", "UTF-8");
```

## 5.3 Attacks by sending HTTP POST requests through WebView.postUrl

The method postUrl loads the URL with postData using"POST" method into the WebView. The method takes two parameters, the URL of the resource to load and the data to be passed to "POST" request.

```
String postData = "recip=user1
&title=WebViewAttackTitle
&message=HttpAttackMessage";
```

```
String attackUrl =
"http://www.trustedSite.com/
cgi-bin/Forum/new_pm.php";
```

```
webView.postUrl(attackUrl,     EncodingUtils.getBytes
(postData, "base64"));
```

The EncodingUtils class has a collection of utility methods for various encoding tasks. The getBytes method converts string to a base64 byte array. The byte array is sent as HTTP POST request to the target site, where the user is logged in. The method leverages the users credentials and sends the HTTP POST request on the behalf of user. The attacks are shown in Fig 3.

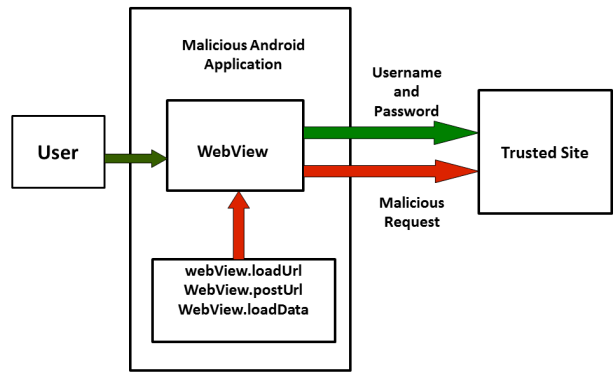

Fig. 3 Attacks using Web-based APIs.

## 5.4 Attacks using HttpClient APIs

HTTP cookies provide the server with a mechanism to store and retrieve state information on the client application's system. This mechanism allows web-based applications, the ability to store information about session IDs, selected items, user preferences, registration information, and other information that can be retrieved later. Android applications can monitor the events occurred within WebView, by overriding the shouldOverrideUrlLoading hook, which is triggered by the navigation event. Cookies can be gathered at every page Navigation of the user using the method getCookie() from
CookieManager class as shown in the code fragment below:

```
CookieManager cookieManager =
CookieManager.getInstance();
```

```
final String cookie =
```

cookieManager.getCookie(url);

The cookies which are gathered can be appended to the list of headers of the HTTP Request using the method setHeader. The attack can be executed by sending the HTTP GET/POST request as shown below:

HttpClient httpClient =
new DefaultHttpClient();

HttpPost httpPost =
new        HttpPost("http://www.targetSite.com/cgi-bin/Forum/new_pm.php");

httpPost.setHeader("Cookie", cookie);

HttpResponse response =
httpClient.execute(httpPost);

The attacker is now able to exploit the user's login session by appending the cookies to the HTTP Request methods, and will be able to send the requests on users behalf resulting in Cross site request forgery.

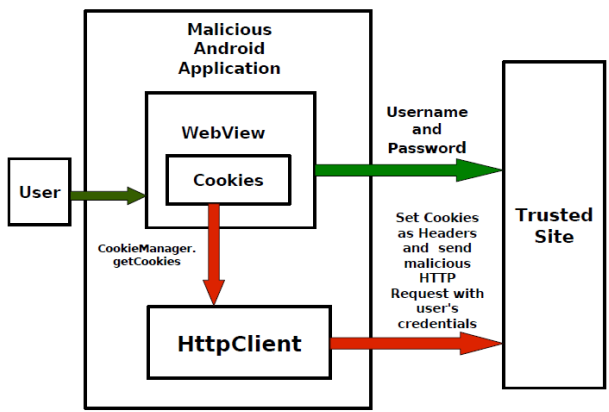

Fig. 4 Attacks using HTTP Client APIs.

## 6. Implementation

To demonstrate the feasibility of such attacks, a simple PHP forum script called Forum installed at the server side [21]. The script uses a backend MySQL database. It has an internal registering system, where users can register their login ID and password; they can post topics and reply to others. At the device side, for the experiments, I used the Celkon's CT2 tab with Android version 4.3.3 (codename Icecream sandwich). A malicious application was developed, for each of the use cases described in section 5. Throughout the paper it was assumed that the user is logged into the malicious application using his username and password. For the attacks related to the Web-based APIs that were described in section 5 automatic form submissions were tested using both GET and POST requests using WebView.loadUrl and WebView.loadData. The form submissions resulted in posting topics by the attacker on behalf of the user. The automatic form submissions using POST requests were also tested using WebView.postUrl API. Attacks were also modeled using the HTTPClient APIs. A malicious application was developed to gather cookies from WebView, by using the CookieManager.getCookie() method. The cookies were set as HTTP headers. The HTTP requests contained session ID that was gathered from cookies. The attacks were tested by sending both GET and POST requests to the Forum site. This resulted in posting topics by the attacker on the Forum site, on behalf of user.

## 6. Conclusion

In the present work, Cross site request forgery attacks on Android Webview is studied. When the user is logged into the trusted site through WebView, the site authenticates the WebView and not the application. The Android application can exploit the user's credentials through the Web-based APIs of WebView. The web-based APIs can submit forms automatically and post malicious content on the trusted site on the behalf of attacker. Three such attacks are modeled using APIs WebView.loadUrl, WebView.loadData and WebView.postUrl. The attacks can also be launched by appending cookies to the HTTP headers and HTTP Requests can be sent to the trusted site using user's credentials. The attacks are easy to execute, but difficult to detect and prevent, as the user is not aware that the attacks are being executed using his credentials. Future work will focus on developing solutions to defend against such attacks on WebView.

**Bhavani A B** received M.Tech degree in VLSI and Embedded Systems, from International Institute of Information Technology, Hyderabad, India in the year 2007.She has worked in the Mobile industry in areas related to Embedded Systems, Mobile technologies, WebKit browser development, Linux Kernel programming, BREW MP and Android. She has the experience in handling the responsibility of the entire project end to end including understanding the requirements and scope, design, coding, integration, debugging, documentation and release. Her research interests include Embedded Systems, Mobile technologies, Digital signal processing, and FPGA Design, EDA and computer security. Right now she is working as Senior Engineer in development of DSP tools at one of the reputed FPGA giants.